\documentclass[showpacs,aps,prd]{revtex4}
\pdfoutput=1

\usepackage{graphicx}
\usepackage{amsmath}

\input{babarsym}
\newcommand{\model}{\ensuremath{\mathrm{(mod)}}\xspace}
\newcommand{\Br}{\BR}
\newcommand{\gevccs}{\ensuremath{{\mathrm{\,Ge\kern -0.1em V^2\!/}c^4}}\xspace}
\newcommand{\ct}{\ensuremath{\cos\theta}\xspace}

\newcommand{\bei}{\begin{itemize}}
\newcommand{\eei}{\end{itemize}}
\newcommand{\beq}{\begin{equation}}
\newcommand{\eeq}{\end{equation}}
\newcommand{\beqn}{\begin{eqnarray}}
\newcommand{\eeqn}{\end{eqnarray}}
\newcommand{\beqns}{\begin{eqnarray*}}
\newcommand{\eeqns}{\end{eqnarray*}}

\def\exp{{\rm exp}}

\def\min{{\rm min}}

\def\rPTbarkappa{\kern 0.18em\overline{\kern -0.18em r}{}^{\kappa}{}}
\def\rPTbarsigma{\kern 0.18em\overline{\kern -0.18em r}{}^{\sigma}{}}

\def\deltabarkappa{\kern 0.18em\overline{\kern -0.18em \delta}{}_r^{\kappa}}
\def\deltabarsigma{\kern 0.18em\overline{\kern -0.18em \delta}{}_r^{\sigma}}
\def\deltaTbarkappa{\kern 0.18em\overline{\kern -0.18em \delta}{}_T^{\kappa}}
\def\deltaTbarsigma{\kern 0.18em\overline{\kern -0.18em \delta}{}_T^{\sigma}}

\def\OC{X}
\def\OCbar{{\kern 0.18em\overline{\kern -0.18em \OC}}}

\def\de{\DeltaE}

\def\a{\kappa}

\def\Amptpbar{\kern 0.18em\overline{\kern -0.18em {\cal A}}_{{\overline B^0} \rightarrow K^-\pi^+\pi^0}}

\def\Amptpbarkappa{\kern 0.18em\overline{\kern -0.18em A}{}^{\kappa}{}}
\def\Amptpbarsigma{\kern 0.18em\overline{\kern -0.18em A}{}^{\sigma}{}}

\def\Tbarkappa{\kern 0.18em\overline{\kern -0.18em T}{}^{\kappa}{}}
\def\Tbarsigma{\kern 0.18em\overline{\kern -0.18em T}{}^{\sigma}{}}
\def\Pbarkappa{\kern 0.18em\overline{\kern -0.18em P}{}^{\kappa}{}}
\def\Pbarsigma{\kern 0.18em\overline{\kern -0.18em P}{}^{\sigma}{}}

\def\Nbpm{{\kern 0.18em\overline{\kern -0.18em N}}^{+-}}
\def\Nbmp{{\kern 0.18em\overline{\kern -0.18em N}}^{-+}}

\def\Mu{\mu}
\def\Chi2MinaMu{\chi^2_{\min ;\a,\Mu}}
\def\Chi2MinMu{\chi^2_{\min ;\Mu}(a)}

\def\fscfave{\kern 0.18em\overline{\kern -0.18em f}_{\rm SCF}}

\def\abar{\bar{a}}

\def\Bbar{\kern 0.18em\overline{\kern -0.18em B}{}\xspace}

\def\BRpmb{{\cal \kern 0.18em\overline{\kern -0.18em  B}}{}_{\rho\pi}^{+-}}
\def\BRmpb{{\cal \kern 0.18em\overline{\kern -0.18em  B}}{}_{\rho\pi}^{-+}}

\def\BRipmb{{\cal \kern 0.18em\overline{\kern -0.18em  B}}{}_{\rho^+\pi^-}}
\def\BRimpb{{\cal \kern 0.18em\overline{\kern -0.18em  B}}{}_{\rho^-\pi^+}}

\def\Abar{\kern 0.18em\overline{\kern -0.18em A}{}}
\def\abar{\kern 0.18em\overline{\kern -0.18em a}{}}

\providecommand{\BDC}{\ensuremath{B^-\to D^{+}\pi^-\pi^-}\xspace}

\begin{document}

\title{\boldmath Dalitz Plot Analysis of~$\Bm \to \Dp\pim\pim$}


\author{T.~M.~Karbach$^1$, representing the \babar\ Collaboration\\
\small\emph{$^1$Technische Universit\"at Dortmund}}

\begin{abstract}
We present a Dalitz plot analysis of $B^-\to D^+\pim\pim$ decays,
based on a sample of about $383$ million $\FourS \to \BB$ decays
collected by the \babar\ detector at the \pep2\ asymmetric-energy
\B~Factory at SLAC. The analysis has been published previously
in~\cite{prdpaper}. We measure the inclusive branching fraction of the
three-body decay to be $\BR(\B^- \to D^+\pim\pim) = (1.08 \pm 0.03\stat
\pm 0.05\syst) \times 10^{-3}$. We observe the established $D^{*0}_2$
and confirm the existence of $D^{*0}_0$ in their decays to $D^+\pim$,
where the $D^{*0}_2$ and $D^{*0}_0$ are the $2^+$ and $0^+$ $c\bar{u}$
P-wave states, respectively. We measure the masses and widths of
$D^{*0}_2$ and $D^{*0}_0$ to be:
$m_{D^{*}_2}    = (2460.4 \pm 1.2 \pm 1.2 \pm 1.9) \mevcc$,
$\Gamma_{D^*_2} = (  41.8 \pm 2.5 \pm 2.1 \pm 2.0) \mev$,
$m_{D^{*}_0}    = (  2297 \pm 8   \pm 5   \pm 19 ) \mevcc$,
$\Gamma_{D^*_0} = (   273 \pm 12  \pm 17  \pm 45 ) \mev$.
The stated errors reflect the statistical and systematic uncertainties,
and the uncertainty related to the assumed composition of signal events
and the theoretical model.
\end{abstract}

\maketitle


\section{INTRODUCTION}
\label{sec:Introduction}

Orbitally exited states of the $D$ meson, denoted as
$D_J$, provide a unique opportunity to test Heavy Quark
Effective Theory (HQET). There are expected to be
four P-wave states of positive parity with the quantum numbers
$0^+(j=1/2), 1^+ (j=1/2), 1^+ (j=3/2)$ and $2^+ (j=3/2)$, which are
labeled as $D^*_0$, $D_1$, $D^\prime_1$ and $D^*_2$, respectively, where
$j$ is the sum of the spin of the light quark and the angular momentum
$L$. Conservation of parity and angular momentum restricts the final
states and partial waves that are allowed in the decays of the various
$D_J$ mesons. The resonances that decay through a D-wave are expected to
be narrow ($\sim30$ \mevcc) and the resonances that decay through an
S-wave are expected to be wide (a few hundred \mevcc). The $D^*_2$ can
only decay via a D-wave and the $D^*_0$ can only decay via an S-wave.
The $D_1$ and $D^\prime_1$ may decay via S-wave and D-wave.
Fig.~\ref{fig:1} shows the spectroscopy of the D-meson excitations and
expected transitions. The Belle Collaboration has reported the first
observation of the broad $D^*_0$ and $D_1^\prime$ mesons in $B$
decays~\cite{belle-prd}. However, the Particle Data Group~\cite{PDG}
considers that the $J$ and $P$ quantum numbers of the $D^*_0$ and
$D_1^\prime$ states still need confirmation.

In this analysis, we fully reconstruct the decays of $B^- \to D^{+}
\pi^-\pi^-$ final states~\cite{conjugate} and measure the inclusive
branching fraction. Then we perform a Dalitz plot analysis to measure
the exclusive branching fractions for $B^- \to D_J^0 \pim$ and to study
the properties of the $D_J$ mesons. The decay $B^-\to D^{+} \pi^-\pi^-$
is expected to be dominated by the intermediate states $D^{*}_2\pi^-$
and $D_0^{*}\pi^-$ and has a possible contribution from the $\Bm\to
D^{+} \pi^-\pi^-$ non-resonant (NR) decay. Also, the $D^*(2007)^0$
(labeled as $D^*_v$) may contribute as a virtual intermediate state, as
well as the $B^{*}$ (labeled as $B^*_v$), produced in a virtual process
$B^-\to B^*\pi^-$. The data used in this analysis were collected with
the \babar\ detector at the \pep2\ asymmetric \epem\ storage rings. The
\babar\ detector is described in~\cite{prdpaper} and in the references
therein. The sample consists of 347.23\,\invfb\ corresponding to
$382.9\pm4.2$ million $\BB$ pairs taken on the peak of the  $\FourS$
resonance. 

\section{EVENT SELECTION}
\label{sec:EventSelection}

\begin{figure}[ht]
  \includegraphics[width=.40\textwidth]{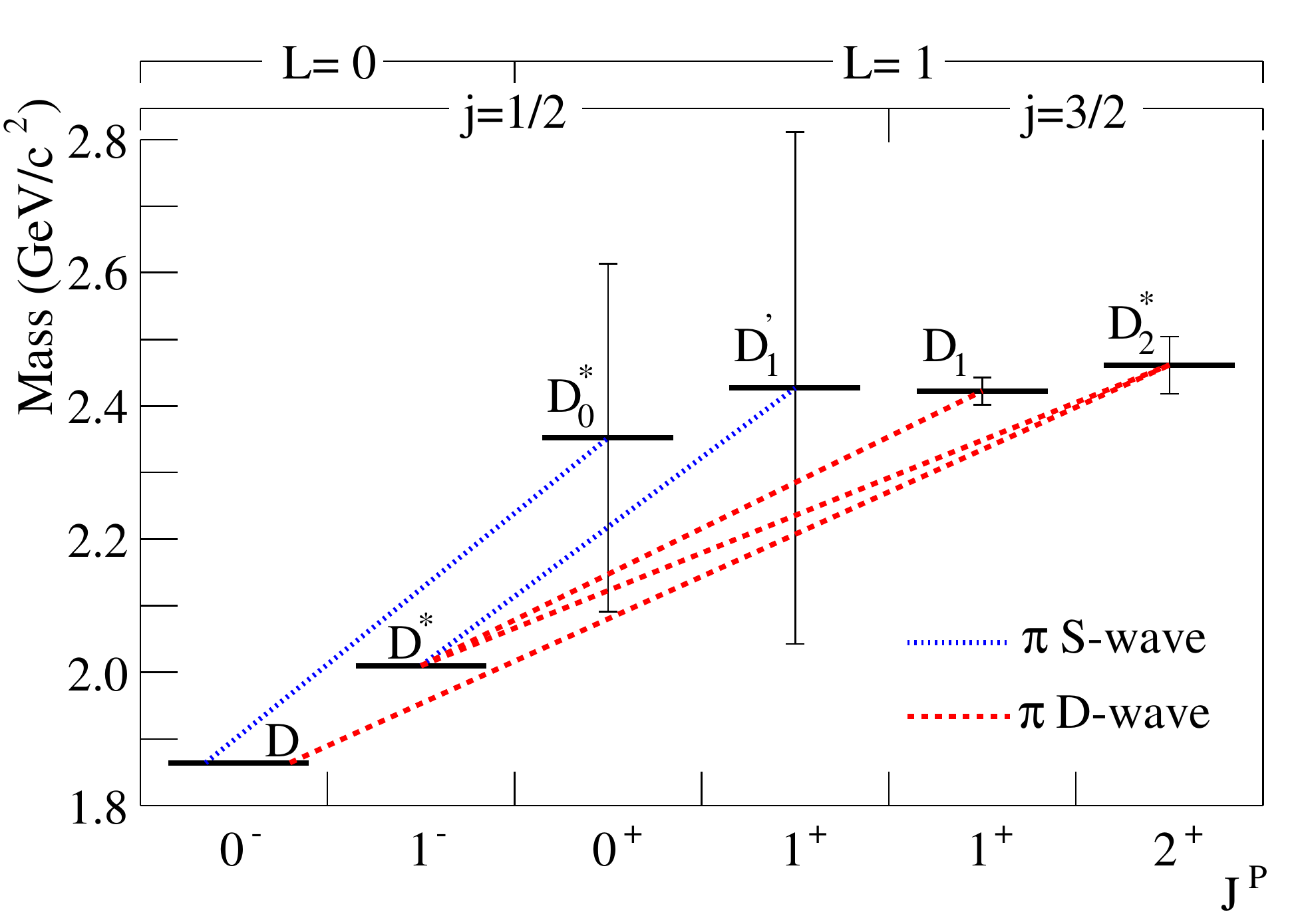}
  \includegraphics[width=.40\textwidth]{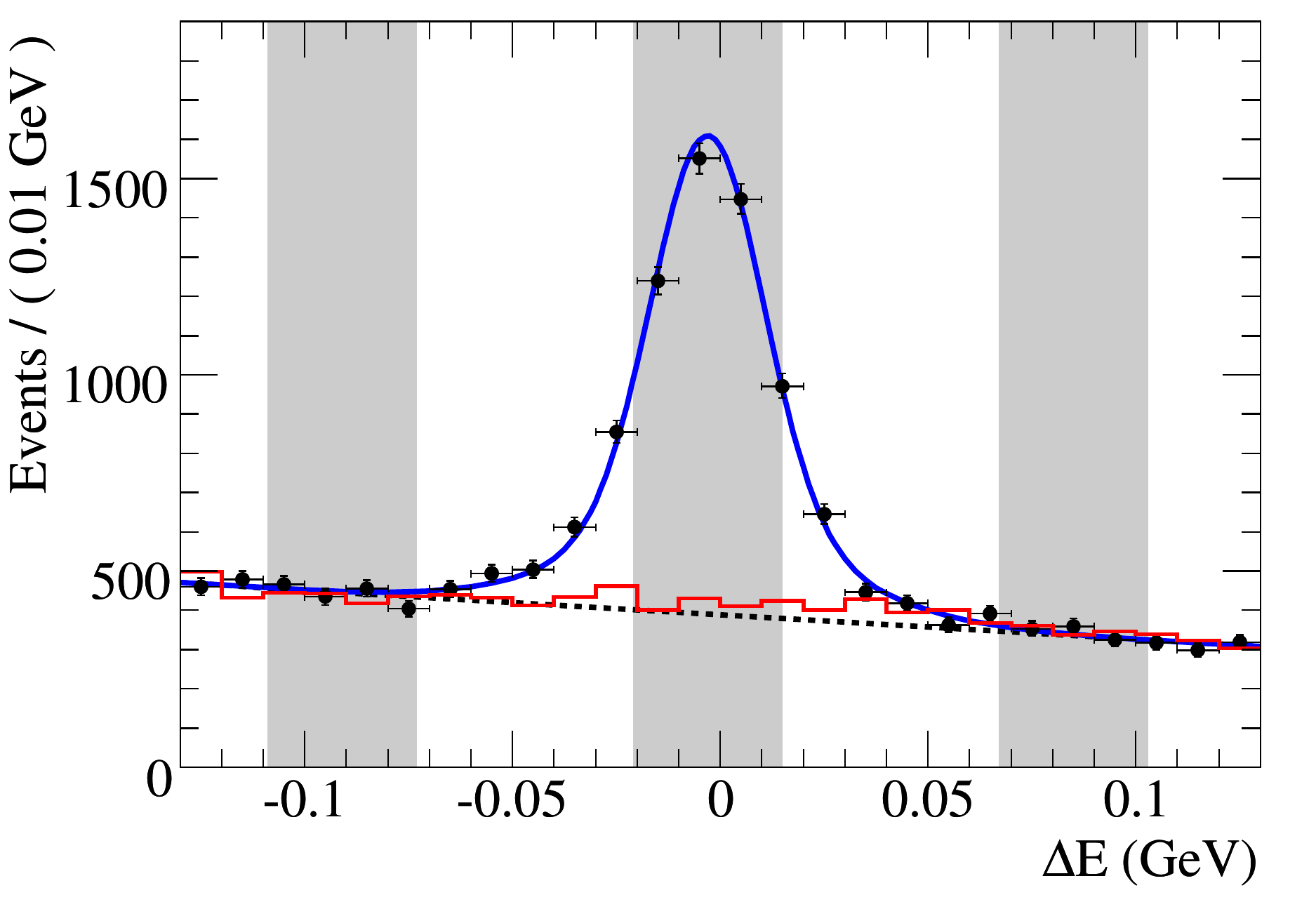}
  \caption{Left: Mass spectrum~\cite{PDG} of $c\overline{u}$ states. The vertical bars
  show the widths. The dotted and dashed lines between the levels show
  the dominant pion transitions. Although it is not indicated in the
  figure, the two $1^+$ states may be mixtures of $j=1/2$ and $j=3/2$,
  and $D^\prime_1$ may decay via a D-wave and $D_1$ may decay via an
  S-wave.
  Right: $\Delta E$ distribution.
  The points with error bars are data, the red curve is the full fit,
  the blue dashed curve is the background, the three shaded regions,
  correspond to the $\Delta E$ left side- and signal bands. The histogram
  shown background expected from MC.}
  \label{fig:1}
\end{figure}

Five charged particles are selected to reconstruct $\Bm\to D^+\pim\pim$
decays with $D^+\to K^-\pi^+\pi^+$. At the $\FourS$
resonance, $\B$ mesons can be characterized by two nearly independent
kinematic variables, the beam energy substituted mass \mes and the
energy difference \de:
\begin{equation}
\label{eq:mesde}
\mes=\sqrt{(s/2+\vec{p_0} \cdot \vec{p_B})^2/E_0^2-p_B^2}, \quad \Delta E=E_B^*-\sqrt{s}/2,
\end{equation}
where $E$ and $p$ are energy and momentum, the subscripts 0 and $B$
refer to the \epem-beam system and the \B candidate respectively; $s$ is
the square of the center-of-mass energy and the asterisk labels the CM
frame. For $\Bm\to D^+\pim\pim$ signal decays, the \mes distribution is
well described by a Gaussian resolution function with a width of
$2.6\mevcc$ centered at the \Bm mass, while the \de distribution can be
represented by a sum of two Gaussian functions with a common mean near
zero and different widths with a combined RMS of $20\mev$. The \de
distribution is shown in Fig.~\ref{fig:1}.

Continuum events are the dominant background. We suppress this
background by restricting two topological variables: the magnitude of
the cosine of the thrust angle, $\cos\theta_{th}$, defined as the angle
between the thrust axis of the selected \B candidate and the thrust axis
of the rest of the event; and the ratio of the second to zeroth
Fox-Wolfram moment~\cite{r2}, $R_2$. Small values of $R_2$ indicate a more
spherical event shape (typical for \BB events) while larger values
indicate a 2-jet event topology (typical for \qqbar events). We also
place restrictions on \mes and \de. Then we fit the \de distribution to
determine the fractions of signal and background events in the selected
data sample. The result of the fit is shown in Fig.~\ref{fig:1}, it
yields $3496\pm 74$ signal events.

To distinguish signal and background in the Dalitz plot studies, we
divide the candidates into three subsamples: the \de signal region, and
two \de sidebands, all defined in Fig.~\ref{fig:1}. A
background MC sample of resonant and continuum events is shown as the
histogram in Fig.~\ref{fig:1}. There is a small amount of
peaking, a fit yields $82\pm 41$ peaking events. The background
subtracted number of signal events is $N_{\rm sig}=3414\pm 85$,
resulting in the background fraction of $(30.4\pm 1.1)\,\%$.

\section{Dalitz Plot Analysis}
\label{sec:dalitzPlotAnalysis}

In this analysis we choose the two $D\pi$ invariant mass-squared
combinations $x=m^2(\Dp\pim_1)$ and $y=m^2(\Dp\pim_2)$ as the
independent variables, where the two like-sign pions are randomly
assigned to $x$ and $y$. This has no effect on our analysis since the
likelihood function (described below) is explicitly symmetrized with
respect to interchange of the two identical particles. We describe the
distribution of candidate events in the Dalitz plot in terms of a
probability density function (PDF). The PDF is the sum of signal and
background components and has the form:
\begin{equation}
\label{eq:pdf}
\mbox{PDF}(x,y) = f_{\rm bg}   \frac{B(x,y)}{\int_{\rm DP} B(x,y) \, {\rm d}x{\rm d}y} 
               + (1-f_{\rm bg})\frac{[S(x,y)\otimes {\cal R}]\,\epsilon(x,y)}{\int_{\rm DP} [S(x,y)\otimes {\cal R}]\,\epsilon(x,y) \, {\rm d}x{\rm d}y},
\end{equation}
where $S(x,y)\otimes \cal R$ is the signal term convolved with the signal
resolution function, $B(x,y)$ is the background term, $f_{\rm bg}$ is
the fraction of background events, and $\epsilon$ is the reconstruction
efficiency. An unbinned maximum likelihood fit to the Dalitz plot is
performed in order to maximize the value of
$\mathcal{L} = \prod_{i=1}^{N_{\rm event}} \mbox{PDF}(x_i,y_i)$
with respect to the parameters used to describe $S$, where $x_i$ and
$y_i$ are the values of $x$ and $y$ for event $i$, respectively.
It is difficult to find a proper binning at the kinematic boundaries in
the $x$-$y$-plane of the Dalitz plot. For this reason, we choose to
estimate the goodness-of-fit $\chi^2$ in the $\cos\theta$ and $m^2_{\rm
min}$ plane, which is a rectangular representation of the Dalitz plot.
The helicity angle $\theta$ is the angle between the momentum of the
pion from the $B$ decay and that of the pion of the $D\pi$ system in the
$D\pi$ restframe; $m^2_{\rm min}$ is the lesser of $x$ and $y$. 
This analysis uses an isobar model formulation in which the signal
decays are described by a coherent sum of a number of two-body ($D\pi$
system + bachelor pion) amplitudes. The orbital angular momentum
between the $D\pi$ system and the bachelor pion is denoted as $L$.
The total decay matrix element $\mathcal{M}$ is then
given by:
\begin{equation}
	\mathcal{M} = \sum_{L=(0,1,2)} \rho_L \, e^{i\Phi_L} [N_L(x,y) + N_L(y,x)] + \sum_k \rho_k \, e^{i\Phi_k} [A_k(x,y) + A_k(y,x)],
\end{equation}
where the first term represents the S-wave ($L=0$), P-wave ($L=1$) and
D-wave $(L=2)$ non-resonant contributions, the second term stands for the
resonant contributions, the parameters $\rho_k$ and $\Phi_k$ are the
magnitudes and phases of the $k^{\rm th}$ resonance, while $\rho_L$ and
$\Phi_L$ correspond to the magnitudes and phases of the non-resonant
contributions with angular momentum $L$. The functions $N_L(x,y)$ and
$A_k(x,y)$ are the amplitudes of non-resonant and resonant terms,
respectively. The resonant amplitudes $A_k(x,y)$ are expressed as
$A_k(x,y) = R_k(m)\,F_L(\rho'r')\,F_L(qr)\,T_L(p,q,\cos\theta)$, where
$R_k(m)$ is the $k^{\rm th}$ resonance lineshape, $F_L(p'r')$ and
$F_L(qr)$ are the Blatt-Weisskopf barrier factors~\cite{formf},
and $T_L(p,q,\cos\theta)$ gives the angular distribution. The parameter
$m (=\sqrt{x})$ is the invariant mass of the $D\pi$ system. The
parameters $p'$, $p$, $q$ and $\theta$ are functions of $x$ and $y$. The
non-resonant amplitudes $N_L(x,y)$ are similar to $A_k(x,y)$ but do not
contain resonant mass terms. The Blatt-Weisskopf barrier factors depend
on a single parameter, $r'$ or $r$, the radius of the barrier, which we
take to be $1.6\,(\gevc)^-1$, similarly to Ref.~\cite{belle-prd}. The functional
forms of the $F_L$ are given in Ref.~\cite{prdpaper}. For virtual
$D_v^*$ decays, $D_v^*\to\Dp\pim$, and virtual $B_v^*$ production in
$\Bm\to B_v^*\pim$, we use an exponential form factor in place of the
Blatt-Weisskopf barrier factor, as discussed in Ref.~\cite{belle-prd}: $F(z) =
\exp(-(z-z'))$, where $z'=rp_v$ for $D^*_v\to\Dp\pim$ and $z'=r'p_v$ for
$\Bm\to B_v^*\pim$. Here, we set $p_v=0.038\gevc$, which gives the best
fit. The resonance mass term $R_k(m)$ describes the intermediate
resonance. All resonances in this analysis are parameterized with
relativistic Breit-Wigner functions:
\begin{equation}
	R_k(m) = \frac{1}{(m_0^2-m^2) - im_0\Gamma(m)}, \quad \Gamma(m) = \Gamma_0 \left(\frac{q}{q_0}\right)^{2L+1} \left(\frac{m_0}{m}\right)\,F_L^2(qr),
\end{equation}
where $m_0$ and $\Gamma_0$ are the values of the resonance pole mass and
decay width, respectively.
The terms $T_L(p,q,\cos\theta)$ describe the angular distribution of
final state particles and are based on the Zemach tensor
formalism~\cite{Zemach}. The definitions are given in~\cite{prdpaper}.
The signal function is then given by $S(x,y) = |\mathcal{M}|^2$.

In this analysis, the masses of $D_v^*$ and $B_v^*$ are taken from the
world averages~\cite{PDG} while their widths are fixed at $0.1\mev$; the
magnitude $\rho_k$ and phase $\Phi_k$ of the $D_2^*$ amplitude are fixed
to 1 and 0, respectively, while the masses and widths of the $D_J$
resonances and the other magnitudes and phases are free parameters to be
determined in the fit. The effect of varying the masses of $D_v^*$ and
$B_v^*$ between $0.001$ and $0.3\mev$ is negligible compared to the
other model-dependent systematic uncertainties.

The fit fraction for the $k^{\rm th}$ decay mode is defined as the
integral of the resonance decay amplitudes divided by the coherent
matrix element squared for the complete Dalitz plot:
\begin{equation}
\label{eq:fk}
f_k = \frac{\int_{\rm DP} |\rho_k(A_k(x,y)+A_k(y,x))|^2 \, {\rm d}x{\rm d}y}{\int_{\rm DP} |\mathcal{M}|^2 \, {\rm d}x{\rm d}y}.
\end{equation}
The detector has a finite resolution. For the narrow resonance $D_2^*$
with the expected width of about $40\mev$, the signal resolution needs
to be taken into account. We study the resolution on MC simulated
events. We find the resolution to be independent of $\cos\theta$ for
truth-matched events, and we describe it by a sum of two Gaussian
functions with a common mean. The signal resolution for an invariant
mass of the $D\pi$ combination around the $D_2^*$ region is about
$3\mevcc$. There is a small self crossfeed (SCF) component, which varies
from $0.5\,\%$ to $4.0\,\%$ with $\cos\theta$. For this component, also
the resolution varies with $\cos\theta$, which we take into account. We
also check the estimated biases in the fitted parameters due to
uncertainties in the signal resolution functions are small.

The signal term is modified in order to take into account the
particle detection efficiency. Since different regions of
the Dalitz plot correspond to different event topologies, the efficiency
is not expected to be uniform over the Dalitz plot. We determine the
efficiency function, $\epsilon(x,y)$, by fitting twice a large sample of
$\Bm\to\Dp\pim\pim$ MC: before and after the final selection was
applied. The properly normalized ratio of the fit functions gives
$\epsilon(x,y)$. The efficiency is flat in the center of the Dalitz
plot, and drops close to its boundaries.

The background distribution is modeled using an analytic function
describing MC background events. Since we find the Dalitz plot
distributions of \de sideband events in data and in MC to be consistent
within their statistics, we are confident that the MC simulation can
accurately represent the background in the signal region.

\section{Physics RESULTS}
\label{sec:Results}

The total $\Bm\to\Dp\pim\pim$ branching fraction is calculated
using the relation:
$\mathcal{B}=N_{\rm signal}/(\overline{\epsilon}\cdot \mathcal{B}(\Dp)\cdot 2N(\Bp\Bm))$,
where $N_{\rm signal}=3414\pm 85$, $\overline{\epsilon}$ is the average
efficiency, $\Br(\Dp)\equiv\mathcal{B}(\Dp\to\Km\pip\pip)=(9.22\pm
0.21)\,\%$, and the total number of $\Bp\Bm$ events $N(\Bp\Bm)=(197.2\pm
3.1)\times 10^6$. The measured total branching fraction is
$\Br(\Bm\to\Dp\pim\pim)=(1.08\pm 0.03\stat)\times 10^{-3}$.

The Dalitz plot distribution for data is included in~\cite{prdpaper}.
Since the composition of events in the Dalitz plot and their
distributions are not known a priori, we have tried a variety of different
assumptions. In particular, we test the inclusion of various components,
such as the virtual $D_v^*$ and $B_v^*$ as well as S-, P-, and D-wave
modeling of the non-resonant component, in addition to the expected 
components of $D_2^*$, $D_0^*$, and background. We choose as the nominal
fit model the one with the $D_2^*$, $D_0^*$, $D_v^*$, $B_v^*$, and
P-wave non-resonant components considered. It produces the best fit
quality with the smallest number of components. The P-wave non-resonant
component is an addition to the fit model used in the previous
measurement from Belle~\cite{belle-prd}. The detailed fit results are:
$m_{D_2^*} = 2460.4 \pm 1.2$,
$\Gamma_{D_2^*} = 41.8 \pm 2.5$,
$m_{D_0^*} = 2297 \pm 8$,
$\Gamma_{D_0^*} = 273 \pm 12$,
$f_{D_2^*} = 32.2 \pm 1.3$,
$\Phi_{D_2^*} = 0.0$ fixed,
$f_{D_0^*} = 62.8 \pm 2.5$,
$\Phi_{D_0^*} = -2.07 \pm 0.06$,
$f_{D_v^*} = 10.1 \pm 1.4$,
$\Phi_{D_v^*} = 3.00 \pm 0.12$,
$f_{B_v^*} = 4.6 \pm 2.6$,
$\Phi_{B_v^*} = 2.89 \pm 0.21$,
$f_{\rm P-NR} = 5.4 \pm 2.4$,
$\Phi_{\rm P-NR} = -0.89 \pm 0.18$,
$f_{\rm bg} = 30.4$ fixed,
where masses are in units of \mevcc, widths in \mev, fractions in \%,
and angles in radians. All errors are statistical only. The total
$\chi^2$ over degrees of freedom is $220/153$. The details of the other
fit models in question are detailed in Ref.~\cite{prdpaper}.
Ref.~\cite{dbugg} argues for an addition of a $D\pi$ S-wave state near
the $D\pi$ system threshold to the model of the $D\pi\pi$ final state.
We have performed according tests~\cite{prdpaper}, which all yielded
worse fit qualities than the nominal fit.

\begin{figure}[ht]
  \includegraphics[width=.3\textwidth]{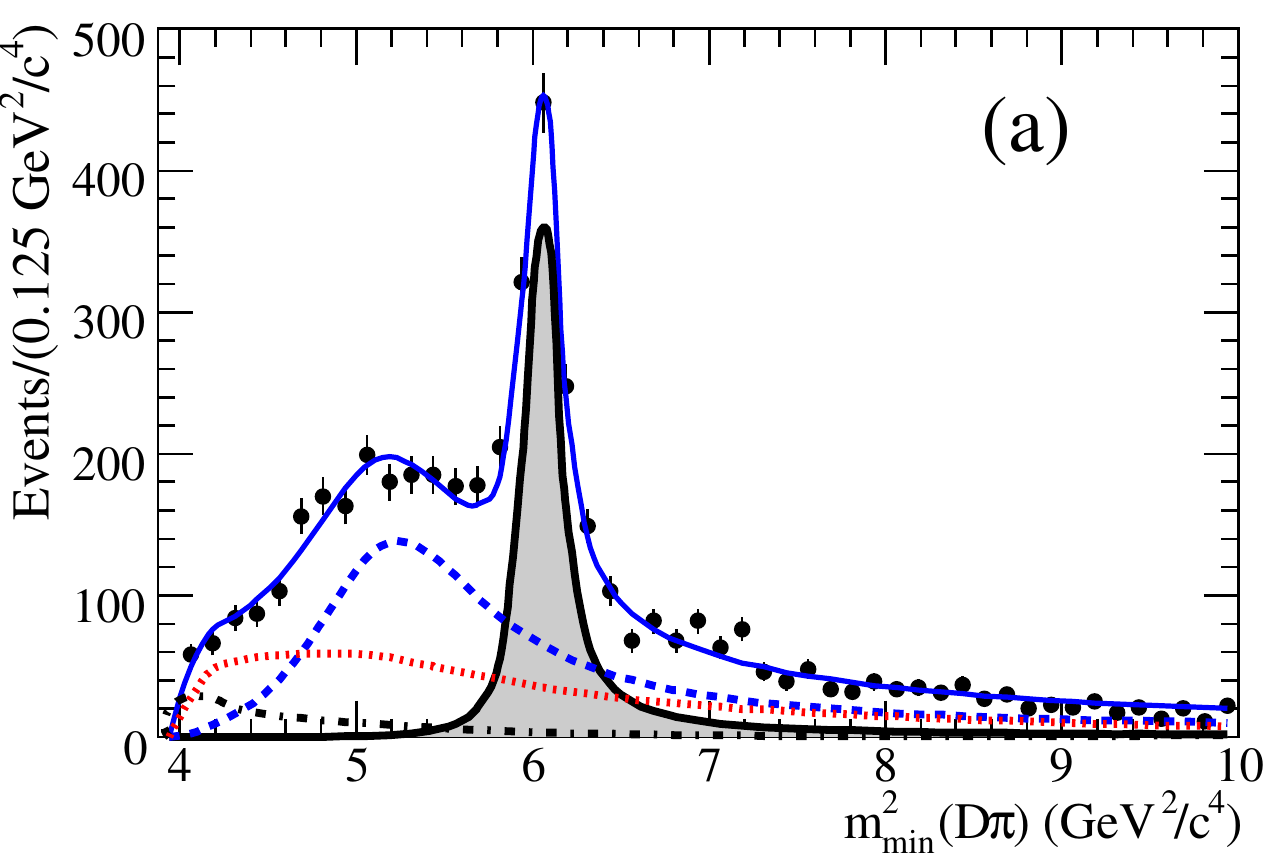}
  \includegraphics[width=.3\textwidth]{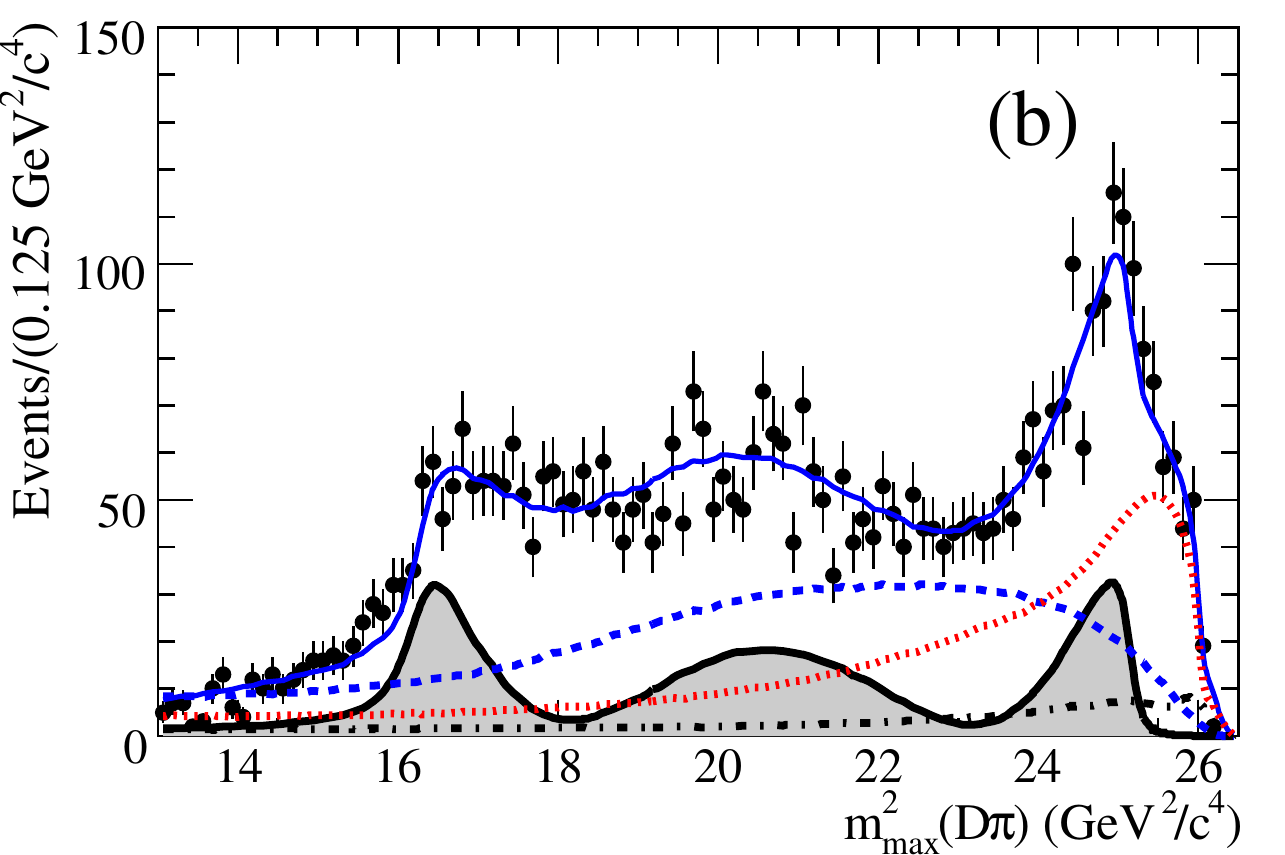}
  \includegraphics[width=.3\textwidth]{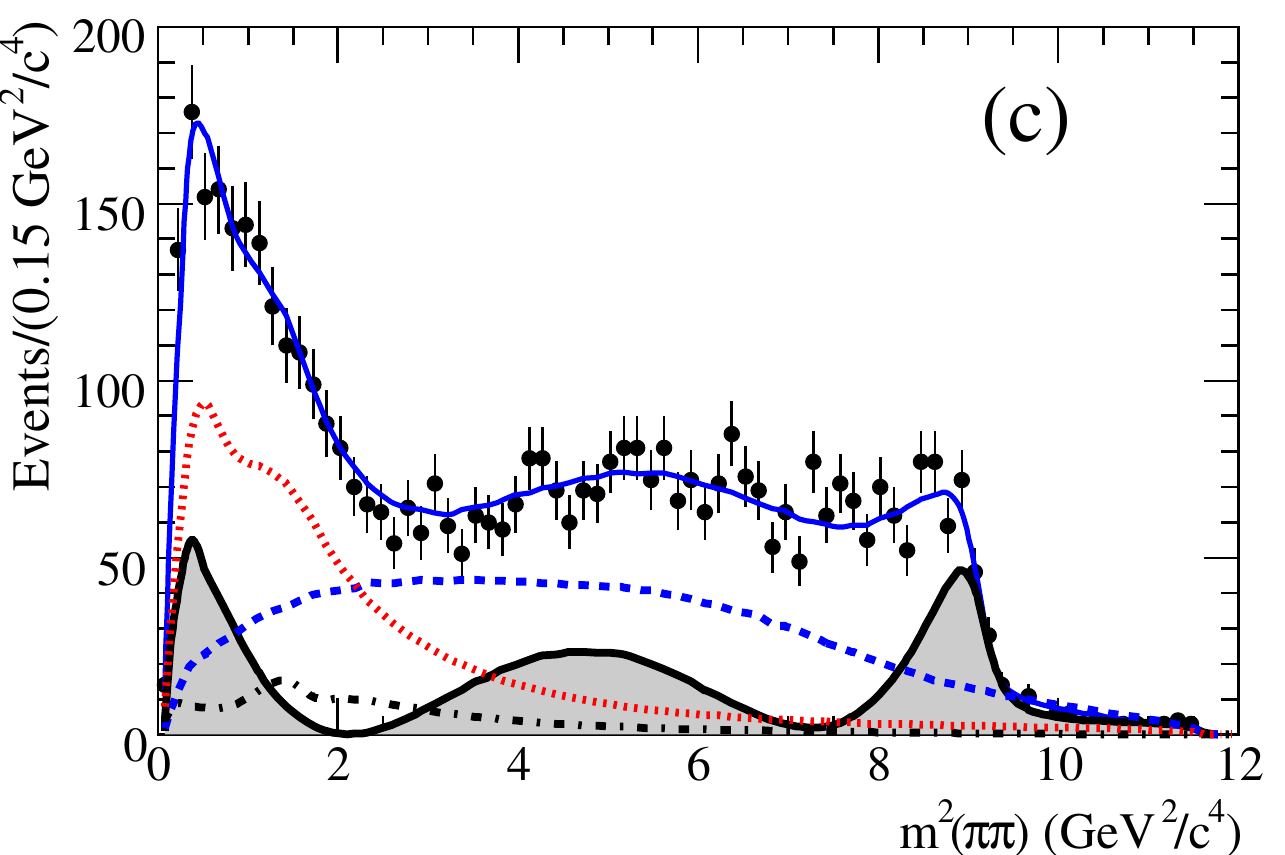}
  \caption{Result of the nominal fit to the data: projections on (a)
  $m_{\rm min}^2(D\pi)$, (b) $m_{\rm max}^2(D\pi)$, and (c)
  $m^2(\pi\pi)$. The points are the data, the solid curves represent the
  nominal fit. The shaded areas show the $D_2^*$ contribution, the
  dashed curves show the $D_0^*$ signal, the dash-dotted curves show the
  $D_v^*$ and $B_v^*$ signals, and the dotted curves show the
  background.}
  \label{fig:dpprojections}
\end{figure}

\begin{figure}[ht]
  \includegraphics[width=.35\textwidth]{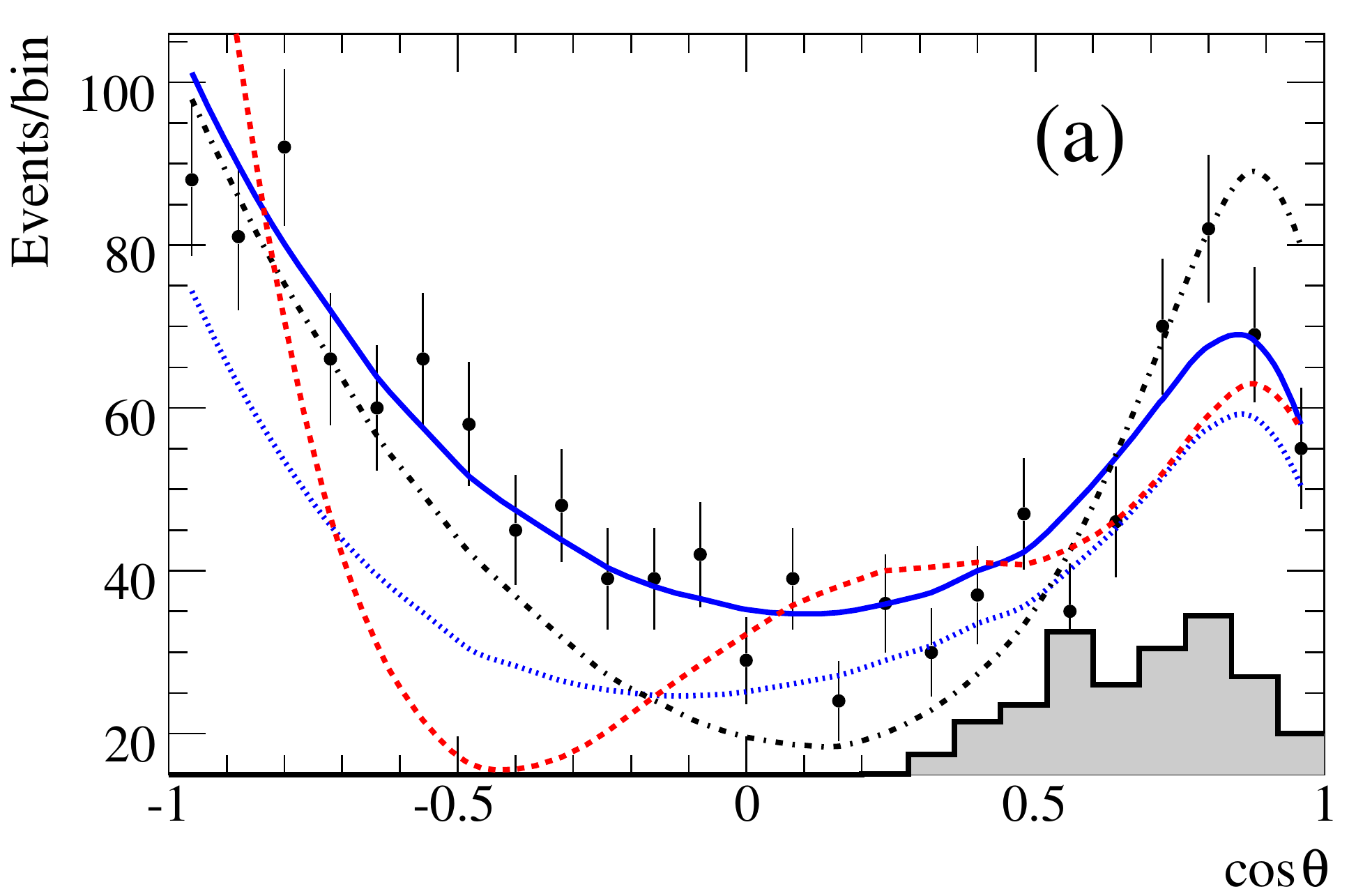}
  \includegraphics[width=.35\textwidth]{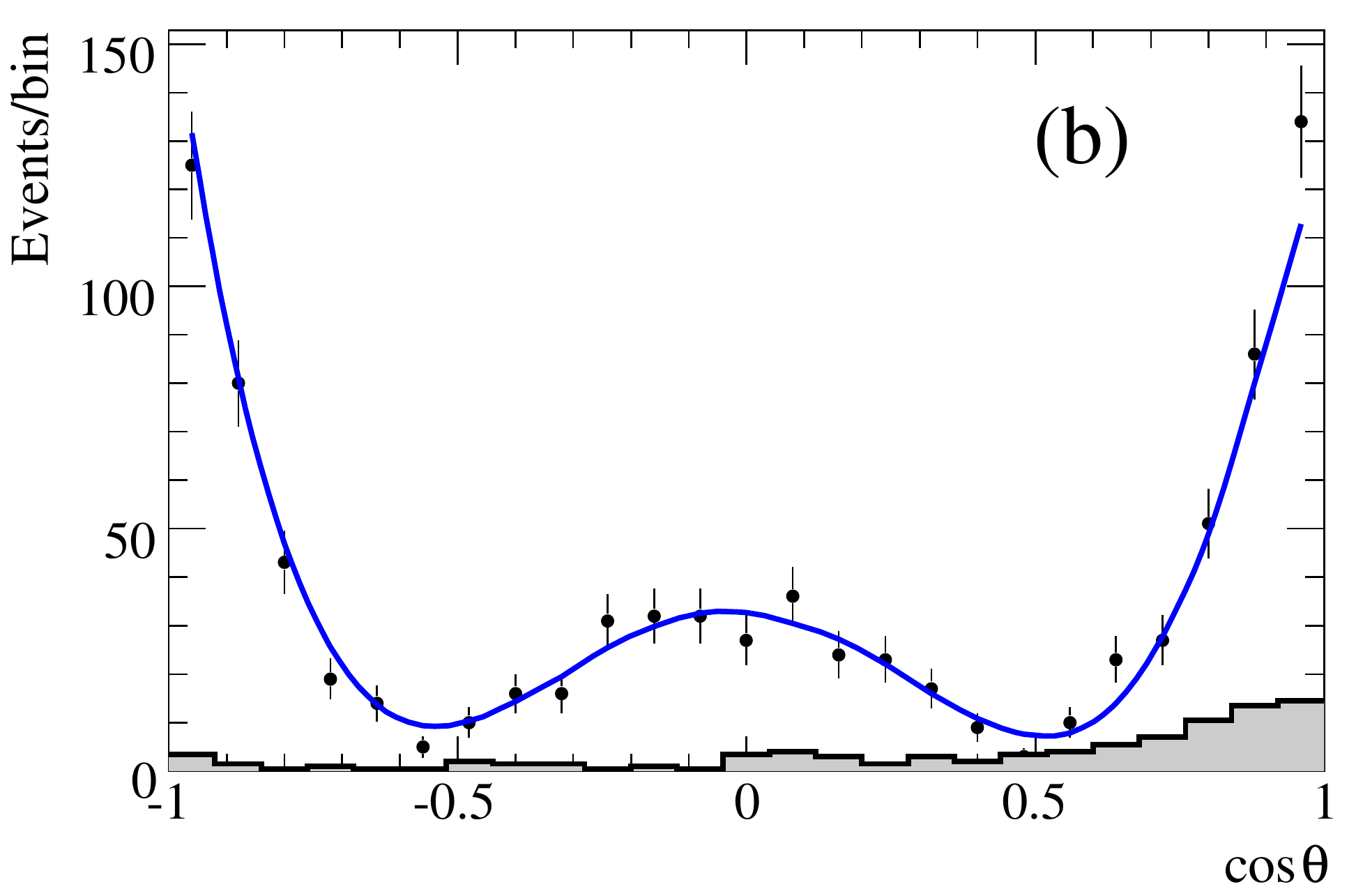}
  \caption{Result of the nominal fit to the data: the $\cos\theta$ distributions
  for  (a) $4.5 < m^2(D\pi) < 5.5 \gevccs$ region and  (b) $5.9 <
  m^2(D\pi) < 6.2 \gevccs$ region. The points with error bars are data,
  the solid curves represent the nominal fit. The dashed, dash-dotted
  and dotted curves in (a) show the fit of hypotheses 2-4 in
  Table~\ref{tab:pwa}, respectively. The shaded histograms show the \ct
  distributions from \de sidebands in data.}
  \label{fig:pwa}
\end{figure}

The nominal fit model results in the following branching fractions:
$\Br(\Bm\to D_2^*\pim)\times\Br(D_2^*\to\Dp\pim)=(3.5 \pm 0.2)\times
10^{-4}$ and $\Br(\Bm\to D_0^*\pim)\times\Br(D_0^*\to\Dp\pim)=(6.8 \pm
0.3)\times 10^{-4}$, where the errors are statistical only.
Fig.~\ref{fig:dpprojections}a-c show the $m_{\rm min}^2(D\pi)$, $m_{\rm
max}^2(D\pi)$, and $m^2(\pi\pi)$ projections, respectively, while
Fig.~\ref{fig:pwa}a and~\ref{fig:pwa}b show the $\cos\theta$
distributions for the $D_0^*$ and $D_2^*$ mass regions, respectively.
The distributions show good agreement between the data and the fit. The
angular distribution in the $D_2^*$ mass region is clearly visible and
is consistent with the expected D-wave distribution of
$|\cos^2\theta-1/3|^2$ for a spin-2 state. In addition, the $D_0^*$
signal and the reflection of $D_2^*$ can be easily distinguished in the
$m_{\rm min}^2(D\pi)$ and $m_{\rm max}^2(D\pi)$ projection,
respectively. The lower edge of $m_{\rm min}^2(D\pi)$ is better
described with the $D_v^*$ component included than without.

Table~\ref{tab:pwa} shows the NLL and $\chi^2/{\rm NDF}$ values for the
nominal fit and for the fits with the broad resonance $D_0^*$ excluded
or with the $J^P$ of the broad resonance replaced by other quantum
numbers. In all cases, the NLL and $\chi^2/{\rm NDF}$ values are
significantly worse than that of the nominal fit. Fig.~\ref{fig:pwa}a
illustrates the helicity distributions in the $D_0^*$ mass region from
hypothesis 2-4; clearly the nominal fit gives the best description of
the data. We conclude that a broad spin-0 state $D_0^*$ is required in
the fit to the data. The same conclusion is obtained when performing the
same tests using the alternative non-nominal fit models.
\begin{table}[htbp]
  \begin{tabular}{clcc}
    \hline
    Hypothesis & Model & NLL & $\chi^2/{\rm NDF}$ \\
    \hline
               & nominal fit & 22970 & $220/153$ \\
    \hline
    1          & $D_2^*$, $D_v^*$, $B_v^*$, P-NR & 23761 & $1171/143$ \\
    2          & $D_2^*$, $D_v^*$, $B_v^*$, P-NR, $(2^+)$ & 23699 & $991/144$ \\
    3          & $D_2^*$, $D_v^*$, $B_v^*$, P-NR, $(1^-)$ & 23427 & $638/135$ \\
    4          & $D_2^*$, $D_v^*$, $B_v^*$, P-NR, S-NR & 23339 & $652/157$ \\
    \hline
  \end{tabular}
  \caption{Comparison of the models with different resonance composition.
  The labels, S-NR and P-NR, denote the S- and P-wave non-resonant contributions.
  \label{tab:pwa}}
\end{table}

The systematic uncertainties under consideration are detailed in
Ref.~\cite{prdpaper}. The systematic effects considered include the
number of $\Bp\Bm$ events, tracking efficiencies, particle
identification, uncertainty on the background shapes, external $\Dp$
branching fraction, and fit bias.

\section{SUMMARY}
\label{sec:Summary}

In conclusion, we measure the total branching fraction of the \BDC decay
to be
$\Br(\Bm \to D^{+}\pim\pim) = (1.08 \pm 0.03\stat \pm 0.05\syst) \times 10^{-3}$.
Analysis of the \BDC Dalitz plot using the isobar model confirms the
existence of a narrow $D_2^*$ and a broad $D_0^*$ resonance as predicted
by HQET. The mass and width of $D_2^*$ are determined to be
$m_{D^*_2} = (2460.4\pm1.2\stat\pm1.2\syst\pm1.9\model) \mevcc$,
$\Gamma_{D^*_2} = (41.8\pm2.5\stat\pm2.1\syst\pm2.0\model) \mev$,
while of the $D_0^*$ they are:
$m_{D^*_0} = (2297\pm8\stat\pm5\syst\pm19\model) \mevcc$,
$\Gamma_{D^*_0} = (273\pm12\stat\pm17\syst\pm45\model) \mev$,
where the third uncertainty is related to the assumed composition of
signal events and the Blatt-Weisskopf barrier factors. The measured
masses and widths of both states are consistent with the world
averages~\cite{PDG} and the predictions of some theoretical models (see
references in~\cite{prdpaper}). We have also obtained exclusive
branching fractions for $D_2^*$ and $D_0^*$ production:
$\Br(\Bm\to D_2^*\pim)\times\Br(D_2^*\to\Dp\pim) = (3.5 \pm 0.2\stat \pm 0.2\syst \pm 0.4\model)\times 10^{-4}$,
$\Br(\Bm\to D_0^*\pim)\times\Br(D_0^*\to\Dp\pim) = (6.8 \pm 0.3\stat \pm 0.4\syst \pm 2.0\model)\times 10^{-4}$.
Our results of the masses, widths and branching fractions are consistent
with but more precise than previous measurements performed by Belle~\cite{belle-prd}.
The relative phase of the scalar and tensor amplitude is measured to be
$\Phi_{D_0^*} = -2.07 \pm 0.06\stat \pm 0.09\syst\pm 0.18\model \, {\rm rad}$.

\section{Acknowledgements}
The author is thankful to the \babar\
Collaboration, its funding agencies, and the organizers of the HADRON09
conference.



\begin{thebibliography}{99}
\bibitem{conjugate} Charged conjugate states are implied throughout the paper.
\end{thebibliography}
\end{document}